\documentclass[%
 reprint,longbibliography,preprintnumbers,
nofootinbib,
 amsmath,amssymb,
 aps,
pre,
]{revtex4-1}
\pdfoutput=1
\usepackage{graphicx}
\usepackage[utf8]{inputenc}
\usepackage{flushend}
\usepackage{dcolumn}
\usepackage{bm}
\usepackage{balance}

\usepackage[normalem]{ulem}

\usepackage[colorlinks = true,
            linkcolor = blue,
            urlcolor  = blue,
            citecolor =green,
            anchorcolor = blue]{hyperref}
\usepackage{verbatim}
\usepackage{color,ulem}
\usepackage[english]{babel}

\usepackage[utf8]{inputenc}

\newcommand{\beq}{\begin{eqnarray}}
\newcommand{\eeq}{\end{eqnarray}}
\usepackage{amsmath}
\usepackage{tikz}
\usetikzlibrary{decorations.pathmorphing}
\usetikzlibrary{shapes.misc}
\tikzset{cross/.style={cross out, draw=black, minimum size=8*(#1-\pgflinewidth), inner sep=0pt, outer sep=0pt},
cross/.default={1pt}}
\usetikzlibrary{patterns,math}
\begin{document}

\title{Theory of sound attenuation in amorphous solids from nonaffine motions}
\author{M. Baggioli}
 \affiliation{Wilczek Quantum Center, School of Physics and Astronomy, Shanghai Jiao Tong University, Shanghai 200240, China.}
 \affiliation{Shanghai Research Center for Quantum Sciences, Shanghai 201315, China.}
\author{A. Zaccone}%
\email{b.matteo@sjtu.edu.cn , alessio.zaccone@unimi.it}
\affiliation{ Department of Physics ``A. Pontremoli'', University of Milan, via Celoria 16,
20133 Milan, Italy
}
\affiliation{Cavendish Laboratory, University of Cambridge, JJ Thomson
Avenue, CB30HE Cambridge, U.K.}

\date{\today}

\begin{abstract}
We present a theoretical derivation of acoustic phonon damping in amorphous solids based on the nonaffine response formalism for the viscoelasticity of amorphous solids. The analytical theory takes into account the nonaffine displacements in transverse waves and is able to predict both the ubiquitous low-energy diffusive damping $\sim k^{2}$, as well as a novel contribution to the Rayleigh damping $\sim k^{4}$ at higher wavevectors and the crossover between the two regimes observed experimentally. The coefficient of the diffusive term is proportional to the microscopic viscous (Langevin-type) damping in particle motion (which arises from anharmonicity), and to the nonaffine correction to the static shear modulus, whereas the Rayleigh damping emerges in the limit of low anharmonicity, consistent with previous observations and macroscopic models.
Importantly, the $k^4$ Rayleigh contribution derived here does not arise from harmonic disorder or elastic heterogeneity effects and it is the dominant mechanism for sound attenuation in amorphous solids as recently suggested by molecular simulations.
\end{abstract}

\maketitle
\section{Introduction}
The mechanism of wave propagation, and, especially, wave attenuation in the disordered states of matter is a central and open topic in contemporary physics, with many ramifications in other fields from plasmas to cosmology.
Since the mid 20th century it is well known that wave attenuation in the acoustic regime of low frequencies and wavelengths is dominated by anharmonic processes at the lowest wavevectors (the so-called \textit{hydrodynamic regime}). In this regime, the sound attenuation, or damping, scales with the wavevector as $\Gamma \sim k^{2}$. This is a diffusive law which can be derived by using conservation equations, i.e. with the methods of hydrodynamic theory and effective field theory~\cite{PhysRevA.6.2401}. \\

In the context of crystals, this is also known as Akhiezer damping~\cite{Akhiezer}, whereas in liquids is known as Brillouin linewidth. The same phenomenon is known to dominate sound attenuation in amorphous solids, such as glasses, at low $k$, where it plays an important role in determining the thermal conductivity~\cite{Ziman}.
More recently, the diffusive nature of vibrational excitations in glasses has been pointed out in numerical simulations~\cite{Allen_1999,Tanaka,Parshin} and used as the starting point for theoretical models of the boson peak caused by anharmonicity in glasses and crystals~\cite{baggioli,PRR}.\\

In glasses \cite{Buchenau_review}, upon going to higher wavevectors, a crossover from $\sim k^2$ to a $\sim k^{4}$ regime is typically observed experimentally~\cite{Masciovecchio}, where the  $\sim k^{4}$  scaling has been interpreted as Rayleigh-type scattering from random fluctuations of some (usually macroscopic) quantity. In this sense, the heterogeneous elasticity theory (HET) has provided a derivation of this Rayleigh type damping based on the assumption of Gaussian spatial fluctuations of the shear modulus~\cite{Schirmacher_2006}. This theory, however, is entirely at the continuum level (its starting point is the elastic modulus, which is a continuum quantity, while the modulus' dependence on microphysics, particle displacements, interactions etc is neglected), hence it does not account for the microsopic structural order/disorder \cite{Milkus} nor for the underlying microscopic (nonaffine) particle dynamics \cite{Lemaitre_PRL_2019,Lemaitre_sound_1,Wei-Ren}.\\

Importantly, it has been recently showed using simulations \cite{szamel2021sound} that the harmonic random fluctuations described by HET are not the only mechanism behind the ubiquitous Rayleigh term $\sim k^4$. Nonaffine motions, which arise from the dynamics in non-centrosymmetric environments \cite{Milkus}, contribute as well to the Rayleigh damping and they are indeed dominant with respect to the HET contribution, even at $T=0$ \cite{szamel2021sound}. This recent observation questions the current paradigm for sound attenuation in amorphous solids and calls for a deeper theoretical understanding of the dominant nonaffine origin of it.\\

While the current models of wave damping in amorphous solids are invariably at the continuum level (or at best effective medium theories~\cite{deGiuli}), much progress has been done recently in the microscopic theoretical description of the viscoelastic response of glasses, through the framework known as \textit{nonaffine response formalism} or nonaffine lattice dynamics~\cite{lubensky,Lemaitre1,Lemaitre2,Scossa,Harrowell,Palyulin}. This theory takes into account the fact that, upon deforming a disordered solid, the particles (atoms, molecules) undergo extra displacements on top of those dictated by the external strain, and these extra displacements, which are random, are called \textit{nonaffine motions}. This framework provides quantitatively accurate predictions of the viscoelastic moduli of glasses with no fitting parameters in good agreement with simulations~\cite{Palyulin}.
It is by now recognized that nonaffine motions play a central role in determining the dynamics of glasses at the microscopic level.\\

In this work, we present a theory of acoustic wave attenuation in amorphous solids based on nonaffine motions and we analytically predict the contributions from anharmonicity to the quadratic part of the damping $\Gamma(k)$. We predict the hydrodynamic diffusive damping at low $k$, including the prefactor which is related to important physical parameters such as the Debye frequency, the nonaffine correction to the shear modulus, and the microscopic friction due to anharmonicity. We provide a useful analytical expression for the Rayleigh contribution $\Gamma \sim k^4$ arising from nonaffinity and identified in the simulations of \cite{szamel2021sound} as the dominant effect on sound damping in amorphous systems. Finally, we also predict the experimentally observed~\cite{Masciovecchio} crossover from diffusive damping to Rayleigh damping at higher $k$ and we give an estimate of the critical wave-length in terms of the previously mentioned physical parameters.\\

\section{Preliminaries}
Within standard linear response theory, the time-dependent expectation value of the stress tensor $\sigma^{ij}(t)$ is given by a linear convolution 
\begin{equation}
\langle \sigma^{ij}(t) \rangle  \simeq \int_{-\infty}^{\infty} \chi_{\sigma\sigma}^{ijkl}(t-t')\,\gamma^{kl}(t') + \mathcal{O}(\gamma^{2}).
\end{equation}
with the strain tensor $\gamma^{ij}$, whose kernel is given by the dynamic response function (sometimes also labelled two-points function, correlator or Green function) $\chi_{\sigma\sigma}^{ijkl}(t)$, i.e. the stress auto-correlation function ~\cite{Chaikin}. Neglecting dissipative terms, the zero frequency limit of the Fourier transformed response function is simply the elastic tensor given in terms of the various elastic constants.\\
From now on, we will focus only on the shear response ($ij=xy$) and therefore Latin indices will be omitted to avoid clutter. Upon Fourier transforming, we have:
\begin{equation}
\sigma(\omega) = \chi_{\sigma\sigma}(\omega)\,\gamma(\omega)\,,
\label{response}
\end{equation}
which is valid in the linear regime, to leading order in the external strain $\gamma$. Importantly, in general, $\chi_{\sigma\sigma}(\omega)$ is a complex-valued function whose real and imaginary components encode respectively the reactive and dissipative parts of the response function and it coincides with the complex dynamic modulus used in viscoelasticity theory. The real and imaginary components are related by the Kramers-Kronig relations \cite{PhysRev.104.1760} (imposed by causality) and in a simple viscoelastic system, to leading order in frequency, are given by (see e.g. Ref.~\cite{pipkin2012lectures}):
\begin{equation}\label{lowlow}
    \chi_{\sigma\sigma}(\omega)\,=\,\mu\,+\,i\,\omega\,\eta\,+\,\mathcal{O}\left(\omega^2\right)\,
\end{equation}
where $\mu$ is the static shear modulus and $\eta$ the shear viscosity.\\

Let us now consider a generic operator $\phi$ and its conjugate external field $\delta h$. We define the dynamic response function $\chi_{\phi\phi}$ associated to such operator as:
\begin{equation}
    \delta \langle \phi \rangle (\omega,k)= \chi_{\phi\phi}(\omega,k)\,\delta h(\omega,k)
\end{equation}
where linear response is assumed as well as time/space translational invariance. The response function is complex-valued, $\chi=\chi'+ i \chi''$ and the Kramers-Kronig relation holds:
\begin{equation}
\chi'(\omega)=\mathcal{P} \int_{-\infty}^{\infty}\,\frac{\chi''(\omega')}{\omega'-\omega}\,\frac{d \omega'}{\pi}.
    \end{equation}
    Let us consider the case of transverse phonons and elasticity. Following standard arguments (cfr. Section 7.3.1 in Ref.~\cite{Chaikin}), we obtain:
    \begin{equation}
        \chi_{\textbf{u}_T\textbf{u}_T}(k,\omega)=\frac{1}{-\rho\, \omega^2+\mu\,k^2- i \omega \,\eta \,k^2}\label{eeq1}
    \end{equation}
    which can be generalized to:
    \begin{equation}
        \chi_{\textbf{u}_T\textbf{u}_T}(k,\omega)=\frac{1}{\rho}\,\frac{1}{- \omega^2+\Omega(k)^2- i \omega \,\Gamma(k)}
    \end{equation}
    where:
    \begin{equation}
        \Omega(k)^{2}= v_T^2 k^2+\dots\,,\qquad \Gamma(k)=D\,k^2+\dots
    \end{equation}
    and $v_T^2=\mu/\rho$ and $D=\eta/\rho$ as expected.\\
    Let us go back to \eqref{eeq1} and try to expand it at low frequency. We immediately obtain:
    \begin{equation}
        \chi_{\textbf{u}_T\textbf{u}_T}(k,\omega)=\frac{1}{k^2\,\mu}+ i \frac{\eta\, \omega}{k^2 \,\mu^2}+\dots\label{lowlow1}
    \end{equation}
    At $\omega=0$, this gives the static susceptibility:
     \begin{equation}
        \chi_{\textbf{u}_T\textbf{u}_T}(k,\omega=0)=\frac{1}{k^2\,\mu}\,,
    \end{equation}
    as reported in Eq.6.4.24 in Ref.\cite{Chaikin}.
 
    Now, comparing \eqref{lowlow1} with \eqref{lowlow}, one immediately realizes that the two response functions are related via:
    \begin{equation}\label{bella}
        \chi_{\sigma_T\sigma_T}(k,\omega)= \mu^2\,k^2\,\chi_{\textbf{u}_T\textbf{u}_T}(k,\omega)
    \end{equation}
    which will be explicitly used in the computations which follow (applied to the nonaffine part of the stress and the displacement).

\section{Nonaffine motions} 
In solids, each particle is displaced to a position defined by the macroscopic strain tensor $\mathbf{F}$, according to
$\mathbf{r}_{i} = \mathbf{F}(\gamma) \cdot  \mathbf{r}_{i,0}$.
This position is called the \emph{affine} position.
Due to the structural disorder, since also the nearest neighbours of particle $i$ are being displaced in the local force field of interaction, the net force acting on $i$ in the affine position is not zero, due to the absence of inversion symmetry. This force vector is denoted as $\mathbf{\Xi}_{i}$ and is called the affine force field since it represents the forces that trigger the nonaffine displacements~\cite{Lemaitre1}. It can be shown that the mean squared $\mathbf{\Xi}_{i}$ is proportional to the mean squared nonaffine displacement, where the nonaffine displacement is defined as $\delta \mathbf{r}_{NA}$ in $\mathbf{r}_{i}(\gamma) = \mathbf{F}(\gamma) \cdot  \mathbf{r}_{i,0} +\delta \mathbf{r}_{NA}$, which gives the final position of the particle $i$ in the deformed frame.

An efficient way of representing the nonaffine displacements is as follows, i.e.  by defining~\cite{Lemaitre1}:
\begin{equation}
    \mathbf{r}_{i} = \mathbf{F}(\gamma) \cdot \mathring{\mathbf{r}}_{i}(\gamma)
\end{equation}
where the new variable $\mathring{\mathbf{r}}_{i}$ does the book-keeping of the nonaffine displacements in the undeformed configuration. The ring notation thus indicates that the particle or material point coordinates are measured in the undeformed frame.
In this formalism, we can write the affine force vector as follows~\cite{Lemaitre1}:
\begin{equation}
\mathbf{\Xi}_{i}=-\frac{\partial U}{\partial \mathring{\mathbf{r}}_{i} \partial \gamma}\bigg\rvert_{\gamma \rightarrow 0}.
\label{Xi_1}
\end{equation}

All this can be summarized into an
equation of motion for the displacement $\mathbf{x}_{i}\equiv\mathring{\mathbf{r}}_{i}(t)-\mathring{\mathbf{r}}_{i}(0)$ of a particle $i$ of mass $m$ of the following form~\cite{Lemaitre1}:
\begin{equation}
m\ddot{\mathbf{x}}_{i}+\zeta\dot{\mathbf{x}}_{i}+\mathbf{H}_{ij}\mathbf{x}_{j}=\mathbf{\Xi}_{i}\gamma,
\label{oscillator}
\end{equation}
with inertial, dissipative and interaction force terms, respectively, on the left hand side and the affine-force field on the right side. Here, $\mathbf{H}_{ij}$ is the Hessian matrix that will be defined shortly below. For ease of notation, we used deformed-frame coordinates.
{This equation of motion, Eq.~\eqref{oscillator}, can be derived from a microscopic reversible particle-bath Caldeira-Leggett Hamiltonian:
\begin{equation}
H=\frac{P^2}{2M}+V(P)+\frac{1}{2}\sum_{\alpha=1}^N\left[\frac{p_{\alpha}^2}{m_{\alpha}}+m_{\alpha}\omega_{\alpha}^2\left(X_{\alpha}
-\frac{F_{\alpha}(Q)}{m_{\alpha}\omega_{\alpha}^2}\right)^{2}\right]\label{CLeq}
\end{equation}
where the dynamic variables $\{P,Q\}$ with no subscript refer to the ``particle'' coupled to a bath of harmonic oscillators (the other particles in the system) labeled with subscript $\alpha$ (for a derivation see the Supplementary material of \cite{Palyulin} and Ref.\cite{Cui_atomic}). The particle is dynamically coupled to the $\alpha= 1...N$ bath oscillators via a bilinear coupling with coefficients $c_{\alpha}$ contained in $F_{\alpha}=c_{\alpha} Q$. The bi-linear coupling between the particle and the oscillators serves as a model to represent the long-range anharmonic interactions between the tagged particle and the other particles in the system. Throughout the literature, the Caldeira-Leggett Hamiltonian has also been shown to provide a mapping onto anharmonic force-fields of real molecules in liquids \cite{Rognoni}. Even though the mapping is only partially accurate, it is nevertheless an effective approximate way of accounting for anharmonicity. Importantly, it should also be noted that if the coupling coefficients $c_{\alpha}$ are all identically zero, then also the viscosity of the system is identically zero, since the friction $\zeta$ in the Eq.~\eqref{oscillator} can be shown to vanish identically, and hence also the low-frequency viscosity $\eta = G'' \omega$ will vanish altogether \cite{Lemaitre1}. This is because the friction $\zeta(t)$ is proportional to $\sum_{\alpha} c_{\alpha}^{2}$ as shown in \cite{zwanzig2001nonequilibrium}. Also, for better tractability we restrict our analysis to the Markovian limit, in which $\zeta$ is not a function of time. Finally, in the following, we shall take units in which $m=1$.}

Applying Fourier transformation {$x_i(t)\sim \exp{i \omega t}$ and eigenmode decomposition of the Hessian matrix $\textbf{H}_{ij}$ with respect to the eigenfrequencies $\nu_p$}, the above equation motion leads to the linear response theory, that was developed in ~\cite{Lemaitre1}, and to the following expression analogue of \eqref{response},
\begin{equation}
\begin{aligned}
\sigma(\omega) &= \chi_{\sigma\sigma}(\omega) \,\gamma(\omega)\\
&=\left(\mu_{A} +\sum_{p}\frac{\hat{\Xi}_{p}^{2}}{\omega^{2} -\nu_{p}^{2} -i\omega\,\zeta}\right)\gamma(\omega)
\label{response_lemaitre}
\end{aligned}
\end{equation}
where $\zeta$ is the (Langevin-type) microscopic damping coefficient for particle motion, and $\hat{\Xi}_{p}$ is the projection of the $3N$-dimensional vector $\mathbf{\Xi}$ onto the $3N$-dimensional eigenvector of the Hessian matrix $| \mathbf{p} \rangle$, i.e. in Dirac's bra-ket notation, $\hat{\Xi}_{p} = \langle \mathbf{\Xi} | \mathbf{p} \rangle$. Furthermore, $\mu_{A}$ is the affine shear elastic modulus, and $\nu_{p}$ denotes the $p$-th eigenfrequency of the solid, i.e. $\nu_{p}^{2} = \lambda_{p}$, where $\lambda_{p}$ is the $p$-th eigenvalue associated with eigenvector $| \mathbf{p} \rangle$. We also omitted prefactors with dimension of volume and assumed that particles masses are all equal to one.
The Hessian matrix is defined as 
\begin{equation}
\mathbf{H}_{ij}=\frac{\partial U}{\partial \mathring{\mathbf{r}}_{i} \partial \mathring{\mathbf{r}}_{j}}\bigg\rvert_{\gamma \rightarrow 0} = \frac{\partial U}{\partial \mathbf{r}_{i} \partial \mathbf{r}_{j}}\bigg\rvert_{\mathbf{r}\rightarrow \mathbf{r}_{0}}
\end{equation}
since $\mathring{\mathbf{r}}(\gamma)\rvert_{\gamma \rightarrow 0}=\mathbf{r}_{0}$.
It is a $3N \times 3N$ matrix with $p=1,...3N$ eigenvalues $\lambda_{p}$ and associated eigenvectors $|\mathbf{p}\rangle$.
In \eqref{response_lemaitre} it is clear the existence of two contributions, one coming from affine displacements and encoded in the affine (infinite-frequency) shear modulus $\mu_{A}$, and a second term in bracket which arises from \emph{nonaffine} motions and is controlled by the quantity $\hat{\Xi}_{p}^{2}$, i.e. the square of the affine force field in the basis of the eigenvectors of the Hessian. Also, the real part of this second term  in bracket represents the (negative) nonaffine contribution to the shear modulus, and is often denoted as $\mu_{NA}$ (or, with alternative notation, $G_{NA}$), defined such that
\begin{equation}
\mu = \mu_{A} - \mu_{NA}\label{defmod}
\end{equation}
is the total shear modulus containing both affine and nonaffine contributions.
Importantly, the nonaffine term has potentially a dissipative (damping) component, as we are going to show next. More precisely, the imaginary part of the second term in the bracket of \eqref{response_lemaitre} is non-zero and it gives a finite contribution to the sound attenuation constant coming entirely from nonaffine motion.

Hence, we can identify a susceptibility associated with nonaffine motions, as follows:
\begin{equation}
\chi_{NA}(\omega)=\sum_{p}\frac{\hat{\Xi}_{p}^{2}}{\omega^{2} -\nu_{p}^{2} -i\omega\,\zeta}\,.
\label{response_function}
\end{equation}
where from now on the $\sigma$ labels will be omitted.\\
{Nonaffinity is not necessarily tied to anharmonicity. On the contrary, the atomic-scale friction $\zeta$ and the emergent viscosity associated to it vanish if the Caldeira-Leggett coupling coefficients $c_{\alpha}$ of $F_\alpha=c_{\alpha} Q$ in Eq.\eqref{CLeq}, which mimic the long-range anharmonic interactions in the single particle description, are all identically zero.} As we will see later, the Rayleigh term in the damping does not depend on this coefficient (and therefore on anharmonicity), consistently with previous approaches from harmonic disorder/heterogeneous elasticity theory (HET) \cite{PhysRevB.82.094205}. Nevertheless, and fundamentally, the induced diffusive ($\sim k^2$) contribution in the damping would not be present in the absence of anharmonicity, thus explaining the reason why in the HET harmonic-disorder approaches \cite{Tomaras_Schirmacher,Schirmacher_2006} such a term must be included forcefully ``by hand''.

We can now transform the discrete sum over eigenstates in \eqref{response_function} into a continuous integral over frequency, by introducing the vibrational density of states (VDOS) of the solid~\cite{Lemaitre1}. We thus obtain~\cite{Milkus2}:
\begin{equation}
\chi_{NA}(\omega)=3\,\rho\int_{0}^{\nu_{D}}\frac{g(\nu)\,\xi(\nu)}{\omega^{2} -\nu^{2} -i\, \omega \,\zeta} d\nu
\label{lemaitre}
\end{equation}
where $\rho=N/V$ is the particle density, $\nu_{D}$ is the Debye frequency and, following \cite{Lemaitre1}, we have defined
\begin{equation}
\xi(\nu)=\langle \hat{\Xi}_{p}^{2} \rangle_{\nu_p \in[\nu +d\nu] }
\end{equation}
where the average is performed for all the projections of $\mathbf{\Xi}$ on eigenvectors $|\mathbf{p}\rangle$ with eigenfrequency $\nu_p \in[\nu +d\nu]$. Note that in previous literature $\xi(\nu)$ was denoted as $\Gamma(\nu)$~\cite{Lemaitre1,Scossa}, not to be confused with the sound attenuation constant discussed in this work.
Note that in \eqref{lemaitre} $\omega$ refers to the frequency of the \emph{external} ``mechanical'' source (in linear response language) $x(t)\sim \exp{i \omega t}$, whereas $\nu$ refers to the microscopic, \emph{internal} eigenfrequencies of the solid that arise from diagonalization of the dynamical matrix $\textbf{H}_{ij}$ (cfr. Eq. 42 in \cite{Lemaitre1}).

In Ref.\cite{Scossa} it was shown that,  for amorphous solids
in $d$ space dimensions,
\begin{equation}
 \langle \mathbf{\Xi}|\mathbf{p}\rangle \langle \mathbf{p} | \mathbf{\Xi}\rangle = d\,\kappa\,
R_0^2\, \lambda_{p}\sum_{\alpha}B_{\alpha,xyxy}
\end{equation}
where $\alpha=x,y,z$ and 
$B_{\alpha,xyxy}$ are coefficients that originate from angular averaging over bond orientation vectors. The scaling with $\lambda_{p}$ has been verified in numerical simulations of different disordered and glassy systems in \cite{Milkus2,Palyulin}. Note that $\mu_{NA}=\frac{1}{V}\sum_{p}\frac{\langle \mathbf{\Xi}|\mathbf{p}\rangle \langle \mathbf{p} | \mathbf{\Xi}\rangle}{\lambda_{p}}$~\cite{Lemaitre1,Scossa}.

The above equation therefore implies that $\xi(\nu) \propto \lambda_{p} \propto \nu^{2}$ and consequently
\begin{equation}
\chi_{NA}(\omega)=c \int_{0}^{\nu_{D}}\frac{\nu^{4}}{\omega^{2} -\nu^{2} -i\omega\, \zeta} \,d\nu\label{eqn1}
\end{equation}
where we used the Debye VDOS, $g(\nu) \sim \nu^{2}$, and hid all the pre-factors into an overall parameter $c$ with units Pa$\times s^3$ whose physical interpretation will be clearer in the following. The scaling $\xi(\nu)g(\nu) \sim \nu^{4}$ has been successfully verified in the simulations of \cite{szamel2021sound}.\\

\section{Sound propagation} Now that we have found the stress-stress auto-correlation function, we can use it to obtain the transverse displacement auto-correlation function $\chi_{\textbf{u}_T\textbf{u}_T}(\omega,k)$ which in the rest of the manuscript will be indicated as $C(\omega,k)$, and that can be written as~\cite{Ziman_quantum,baldi,baowen,Chaikin}:
\begin{equation}
    C(\omega,k)\,=\,\frac{1}{m}\,\frac{1}{-z^2+E(k)^{2}- \Sigma(\omega,k)} \label{eq1}
\end{equation}
where $z \equiv \omega+ i \epsilon$ is the complex-valued frequency, $E(k)$ is the bare phonon energy $E(k)^{2}=v^2 k^2$ (with $v$ the Born speed of transverse sound in this case), $\Sigma(\omega,k)$ the self-energy and $m$ the mass density. Whether in the numerator one has $1$ or $2z$ depends on the arbitrary choice of normalization of the bosonic operators in the Hamiltonian \cite{Khomskii}, here we follow the convention of Ziman \cite{Ziman_quantum}.
By setting the self-energy to zero, and solving for the poles of the Green function in \eqref{eq1}, we obtain $\omega=\pm v \,k$, the dispersion relation of the acoustic phonons in which the effects of nonaffinity are neglected. Again, explicit indices are not shown and we consider only the transverse phonons correlation function which corresponds to the shear stress response discussed before. Upon decomposing the self-energy into its real and imaginary parts,
\begin{equation}
    \Sigma(\omega,k)=\Sigma'(\omega,k)+i\, \Sigma''(\omega,k)\,,
\end{equation}
then the dispersion relation of the sound mode can be obtained as a solution of the following equation:
\begin{equation}
    \omega^2= v^2 k^2 - \Sigma'(\omega,k)- i\, \Sigma''(\omega,k)
\end{equation}
which must be compared with the general form:
\begin{equation}
    \omega^2=\Omega(k)^2- i \,\omega\, \Gamma(k).
\end{equation}
where $\Gamma(k)$ is the sound attenuation~\cite{Chaikin}.\\
A simple comparison implies
\begin{equation}
    \Omega(k)^2= v^2 k^2-\Sigma'(v k,k)\,,\qquad  \Gamma(k)=\frac{1}{v k}\Sigma''(v k,k)\,, \label{eq2}
\end{equation}
where we have substituted $\omega \rightarrow v k$ assuming the limit of small wave-vector.\\
In other words, the expressions in \eqref{eq2} are valid only in the region where $\mathrm{Re}\,\omega = v k$. Equation \eqref{eq2} is in perfect agreement with the discussion in \cite{szamel2021sound} and will serve as starting point to derive the sound damping based on the nonaffine formalism in the next sections.
\subsection{The correction to the shear modulus and the nonaffine dynamic response function}
    Finally, as a consistency check of our analysis, we can study the nonaffine correction to the sound speed given by:
\begin{equation}\label{ww}
  v'^2=v^2-\frac{\Sigma'(v k,k)}{k^2}=\frac{\mu_A-\mu_{NA}}{m},
\end{equation}
and check that it correctly reaches a constant value in the limit $k \rightarrow 0$.
For that to be true, we need that:
\begin{equation}
    \lim_{k \rightarrow 0}\Sigma'(v k,k) \sim k^2\,.\label{vcorr}
\end{equation}
Now:
\begin{equation}
    \mathrm{Re}\,C(vk,k)\,=\,-\frac{1}{m}\,\frac{\Sigma'(vk,k)}{\Sigma'(vk,k)^2+\Sigma''(vk,k)^2}
\end{equation}
and then:
\begin{equation}
   \mathrm{Re}\,\chi_{NA}(vk,k)\,\sim -\,\frac{k^2\,\Sigma'(vk,k)}{\Sigma'(vk,k)^2+\Sigma''(vk,k)^2}
\end{equation}
where all the dimensionful parameters are omitted since not relevant for the purpose of this analysis.\\
From \eqref{eqn1}, we see that in the limit of zero frequency (and therefore equivalently momentum), the real part of the nonaffine dynamic response function goes to a negative constant, $\mathrm{Re}\,\chi_{NA}(vk,k)\,\sim\,-\beta$, with $\beta>0$. Therefore, we have that:
\begin{equation}
 \lim_{k \rightarrow 0}\,\frac{k^2\,\Sigma'(vk,k)}{\Sigma'(vk,k)^2+\Sigma''(vk,k)^2} \sim \beta>0
\end{equation}
Now, we know that $\Sigma'' \sim k^3$, in the regime of small frequency/momentum, meaning that:
\begin{equation}
  \lim_{k \rightarrow 0}\frac{k^2\,\Sigma'(vk,k)}{\Sigma'(vk,k)^2+k^6} \sim \beta>0
\end{equation}
and therefore, expanding at small wave-vector, we get $\Sigma' \sim k^2/\beta >0$ which indeed gives a constant, and importantly a \emph{negative} correction to the speed of sound and the static shear modulus (see \eqref{ww}), in agreement with previous numerical calculations~\cite{Milkus2,Palyulin,szamel2021sound} and theoretical analysis~\cite{Lemaitre1,Lemaitre2,Scossa}.
\subsection{The physical meaning of the $c$ parameter}
The previous analysis is helpful to re-write the dimensionful parameter $c$ appearing in the main text in terms of more transparent physical quantities. By restoring all the factors in the above formulas, one obtains that at low energy:
\begin{equation}
    \mathrm{Re}\,\chi_{NA}(v k, k)\,=\,-\,\frac{\mu_{NA}^2\,k^2}{m}\,\frac{1}{\Sigma'(vk,k)}\,.\label{u1}
\end{equation}
Moreover, by using \eqref{ww}, we can rewrite the real part of the self-energy as:
\begin{equation}
    \Sigma'(vk,k)\,=\,\frac{\mu_{NA}}{m}\,k^2\,+\,\dots\label{u2}
\end{equation}
\eqref{u2} together with \eqref{u1} imply:
\begin{equation}
    \mathrm{Re}\,\chi_{NA}(v k, k)\,=\,-\,\mu_{NA}
\end{equation}
which allows us to fix $c$ as:
\begin{equation}\label{bellissima}
    c\,=\,\frac{3\,\mu_{NA}}{\nu_D^3}\,.
\end{equation}
The result above will be useful to express the sound attenuation constant in terms of fundamental physical quantities.
\subsection{Sound attenuation}
By taking the imaginary part of the auto-correlation function in \eqref{eq1}, we obtain:
\begin{equation}
    \mathrm{Im}\,C(v k,k)\,=\frac{1}{m}\, \,\frac{\Sigma''(vk,k)}{\Sigma''(vk,k)^2+ \Sigma'(vk,k)^2}
\end{equation}
where we also used the low-energy expression $\omega=vk$.\\
As demonstrated in \cite{Chaikin} and in \eqref{bella}, a simple relation between the stress auto-correlation function $\chi_{\sigma\sigma}$  and the displacements auto-correlation function $\chi_{uu} $ exists. Restricting it to the nonaffine components, that yields:
\begin{equation}
     \chi^{NA}_{\sigma \sigma}(\omega,k)  = \mu_{NA}^2 k^2  \chi^{NA}_{uu}(\omega,k)  = \mu_{NA}^2\, k^2 C_{NA}(\omega,k).\label{rel}
\end{equation}
This (recall that the affine part of $\chi(\omega)$ is independent of $\omega$~\cite{Lemaitre1}) finally implies:
\begin{equation}
    \mathrm{Im}\,\chi_{NA}(v k,k)\,=\,\frac{\mu_{NA}^2\,k^2}{m}\,\frac{\Sigma''(vk,k)}{\Sigma''(vk,k)^2+ \Sigma'(vk,k)^2}.
\end{equation}
Now, using \eqref{eq2} together with \eqref{defmod}, one can immediately verify that the real part of the self-energy at low momentum is given by $\Sigma'(vk,k)=\mu_{NA}/m \,k^2$. Hence, using $\Sigma'' = v k \Gamma$, we get:
\begin{equation}
    \mathrm{Im}\, \chi_{NA}(v k,k)\,= \,\frac{\Gamma(k)\,  k\, \mu_{NA} ^2 \,m\, v}{k^2 \mu_{NA}^2+\Gamma(k) ^2 m^2
   v^2}.
   \label{connection}
\end{equation}
Now, let us assume that $\Gamma$ scales faster than $k$ (as observed in all experimental and simulations results), therefore at low $k$ we have:
\begin{equation}
    \mathrm{Im}\, \chi_{NA}(v k,k)\,=m\, v\,\frac{\Gamma(k)}{k}\,. \label{ee}
\end{equation}
This clearly implies that a linear in $k$ term in the imaginary part of the self-energy will produce a diffusive $\sim k^{2}$ in the damping, while a term $k^{3}$ will produce the Rayleigh damping $\sim k^{4}$, as we are going now to verify by direct evaluation of the nonaffine theory developed above.

We now recall \eqref{lemaitre} and perform the integral analytically, obtaining:

\begin{align}
\chi_{NA}(\omega)=&c \int_{0}^{\nu_{D}}\frac{\nu^{4}}{\omega^{2} -\nu^{2} -i\,\omega\, \zeta} d\nu = \nonumber\\& -\frac{c\,\nu_D^3}{3} \nonumber- c \,\nu_D\,\omega\,(\omega- i \zeta) \nonumber\\&+c\, \omega^{3/2} (\omega-i \zeta )^{3/2} \tanh
   ^{-1}\left(\frac{\nu_D}{ \sqrt{\omega(\omega-i \zeta )}}\right)\,.
\label{integral}
\end{align}

Let us first discuss the behaviour at the lowest order in frequency.
In that limit, the above integral becomes:
\begin{equation}
\chi_{NA}(\omega)=  -\frac{c\,\nu_D^3}{3}+i \,\zeta \,c \,\nu_D\,\omega+\dots
\label{integral2}
\end{equation}
implying that:
\begin{equation}
    \Gamma(k) =  \,c \,\frac{\zeta\,\nu_D}{m}\,k^2\, .
\end{equation}
This result can be simplified further by using \eqref{bellissima}, leading to:
\begin{equation}
    \Gamma(k) =  \, \,\frac{3\,\zeta\,\mu_{NA}}{m\,\nu_D^2}\,k^2\, . \label{mainresult}
\end{equation}
This is an important result: it shows that the coefficient for diffusive sound damping is proportional to the microscopic Langevin friction $\zeta$, which in turn is related to anharmonicity, and it is also proportional to  $\mu_{NA}$, meaning that in a perfect centrosymmetric crystal at zero temperature (hence with $\mu_{NA}=0$), the sound damping is identically zero. This is consistent with the expectation that sound damping is present only in anharmonic crystals with defects and/or thermal fluctuations (which also cause nonaffinity~\cite{PhysRevE.82.041115,PhysRevE.87.042801,hoover}), and in amorphous solids.

Despite the dependence of the leading quadratic term in Eq.\eqref{mainresult} on the frictional parameter $\zeta$ might appear surprising, it is not. Indeed, it is well known that the diffusive damping in solids is proportional to the viscosity $\eta$ (see for example \cite{Chaikin}), $\Gamma \sim \eta k^2$. The latter grows with the internal friction parameter $\zeta$ leading to $\Gamma \sim \zeta\,k^2$ as derived in Eq.\eqref{mainresult}. From a physical perspective, it is expected that a larger sound attenuation corresponds to larger internal friction, as mathematically derived in Eq.\eqref{mainresult}. \\

{Let us emphasize, that the presence of a diffusive attenuation constant $\Gamma \sim D k^2$ for the collective excitations does not indicate dissipation of energy and/or momentum which are perfectly conserved quantities in the full system. This term is ``dissipative'' in the sense that it breaks time-reversal invariance (reversibility) and it is connected with entropy production. This is consistent with the discussion that the Reader can find in \cite{doi:10.1021/jp503647s}.}

Hence, we find a contribution from nonaffine motions to the ubiquitous \emph{diffusive} damping, which has been observed in countless experimental and simulation studies of glasses, liquids and supercooled liquids~\cite{Ruffle,Tanaka,Hansen}.
Importantly, the microscopic theory explains that the nonaffine contribution to the diffusion coefficient $D$ in $\Gamma_{\textrm{diff}} = D k^{2}$ is proportional to the microscopic damping coefficient for particle motion, $D \propto \zeta$, which clarifies the close connection of the diffusive damping with anharmonicity of particle motion. The Langevin-type damping $\zeta$ can be shown, through particle-bath models of the Caldeira-Leggett type~\cite{Zwanzig1973}, to arise from long-range interparticle interactions, which in condensed matter states are due to the long-range anharmonic part of the interparticle potential~\cite{Palyulin,Rognoni}.
The most general derivation yields a non-Markovian friction coefficient $\zeta(t)$ which, however, reduces to a Markovian, time-independent coefficient $\zeta$, just as the one used here, when the coupling coefficients between each particle and the other oscillators are all the same~\cite{Zwanzig1973,Zwanzig}. This suffices for the sake of our analysis and the more complex non-Markovian case is left for future studies.\\

In general, the expansion of the expression in \eqref{integral} is complicated and not particularly illuminating, nevertheless it becomes rather simple in the limit of small microscopic friction, $\zeta\ll 1$. The imaginary part of the $\tanh
   ^{-1}$ in the regime of $\omega \gg \zeta$ and $\omega \ll \nu_D$ (small friction and low frequency) is approximately constant and equal to $-\pi/2$, leading to:
\begin{equation}
\begin{split}
\mathrm{Im}\chi_{NA}(\omega)=  c\,\nu_D\,\omega \,\zeta+\,c\, \frac{\pi}{2}\,\omega^3+\dots
 \end{split}
\label{appr}
\end{equation}
This result implies a smooth crossover between a diffusive damping at low wave-vector, $\Gamma(k) \sim k^{2}$, and a Rayleigh one, $\Gamma(k)\sim k^4$, at larger wave-vector:
\begin{equation}
    \Gamma(k)=\Gamma_2 k^2 + \Gamma_4 k^4\,.\label{crosso}
\end{equation}
Importantly, the weight of the diffusive term at low wave-vector is controlled by the size of the microscopic Langevin friction $\zeta$:
\begin{equation}
    \frac{\Gamma_2}{\Gamma_4}= \frac{2}{\pi}\frac{\nu_D\,\zeta}{v^2}
\end{equation}
and it vanishes in absence of microscopic friction, $\zeta=0$.

\begin{figure}[ht]
    \centering
    \includegraphics[width=\linewidth]{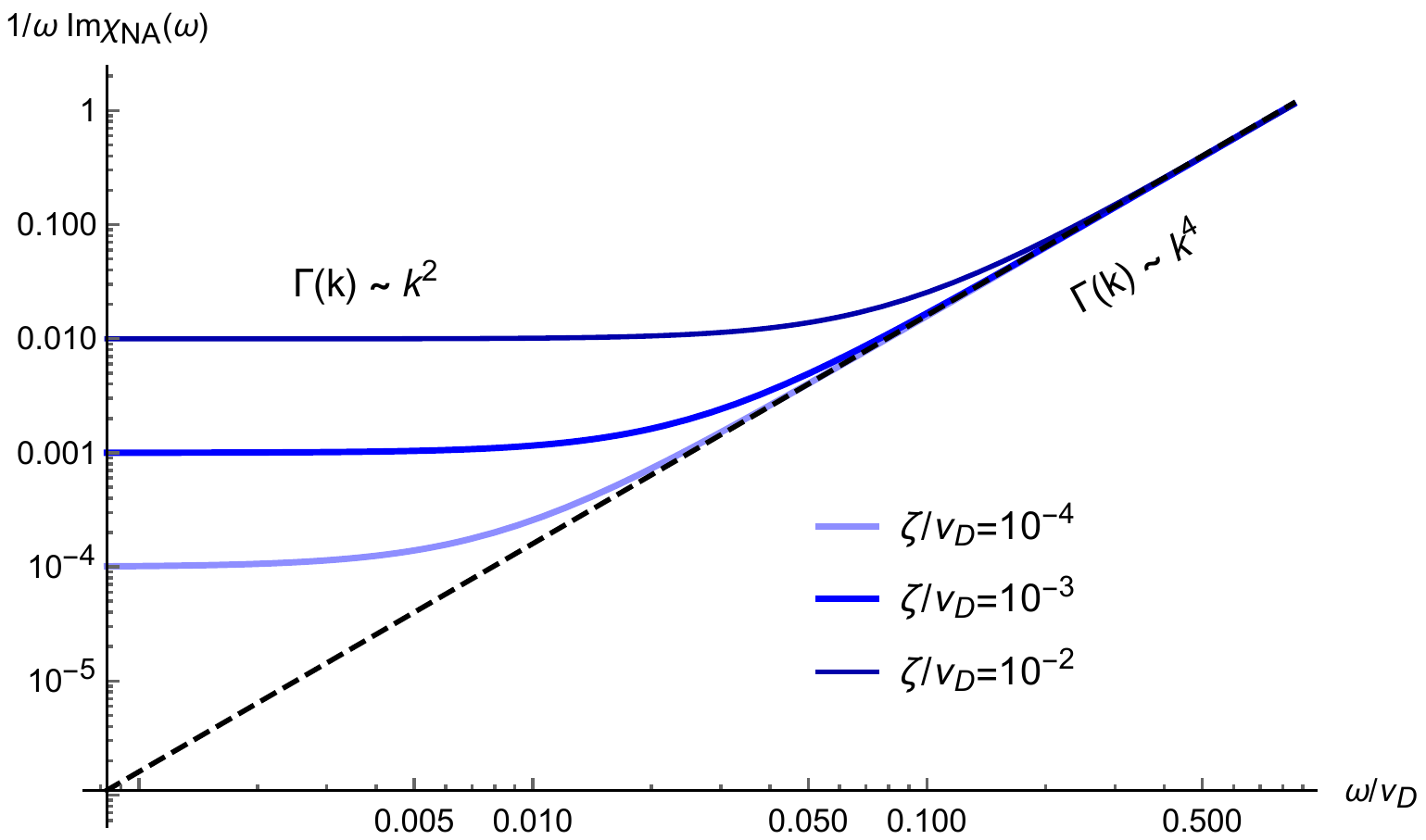}
    \caption{The imaginary part of the nonaffine stress autocorrelation function as a function of the frequency $\omega$. The different colors represent different values of the microscopic damping for particle motion, $\zeta$. The dashed line guides the eye towards the Rayleigh damping scaling. The different regimes of sound damping $\Gamma(k)$ predicted by the theory (via Eq.\eqref{connection}) are indicated. The dimensionful parameter $c$ is set to unity.}
    \label{fig:2}
\end{figure}

The behaviour of the imaginary part of the nonaffine stress correlation function is shown in Fig.~\ref{fig:2} for different strengths of the microscopic damping $\zeta$. The full function displays a smooth crossover from a diffusive regime ($\mathrm{Im}\chi\sim \omega$) to a Rayleigh regime ($\mathrm{Im}\chi\sim \omega^3$), whose location is controlled by the value of  $\zeta$. The larger the $\zeta$, the more extended (and therefore more important) the diffusive regime. At very low values of the microscopic damping, the function is well approximated by \eqref{appr} and the diffusive regime is pushed to very low frequency, whereas the Rayleigh one becomes more predominant.\\

\begin{figure}[ht]
     \centering
   \includegraphics[width=0.8\linewidth]{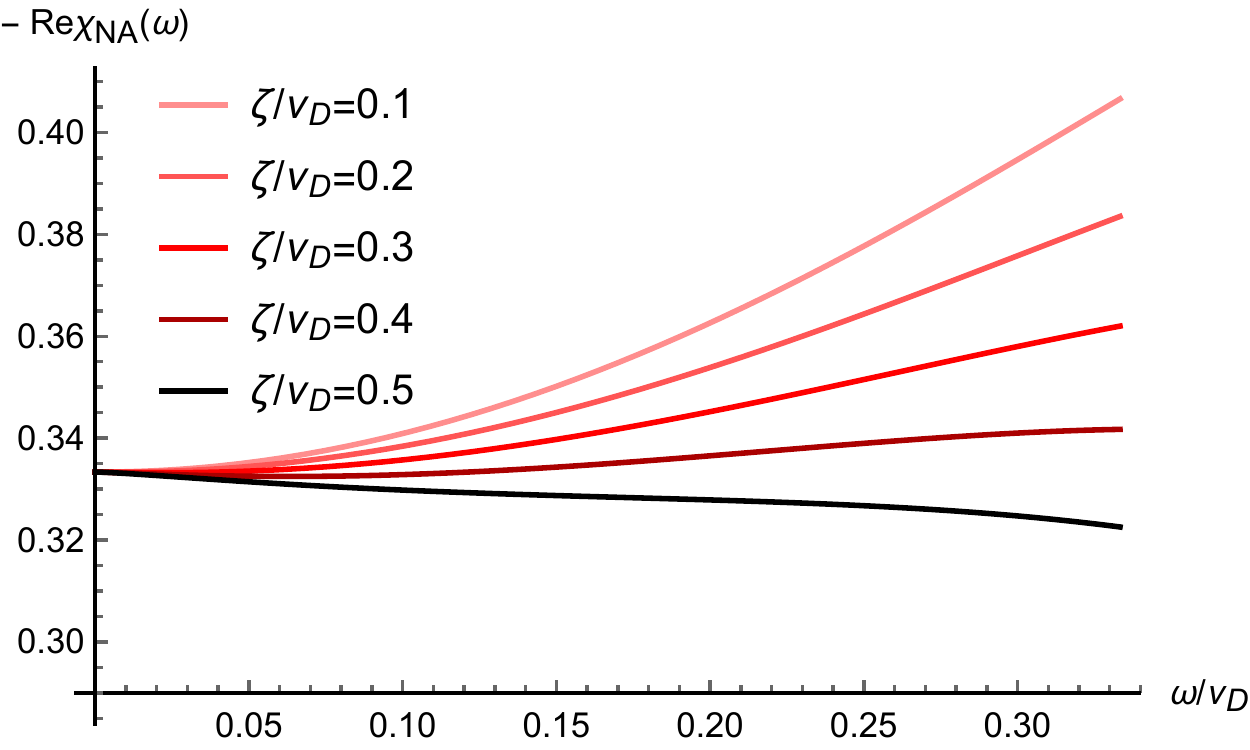}
     
    \vspace{0.4cm}
     
      \includegraphics[width=0.8\linewidth]{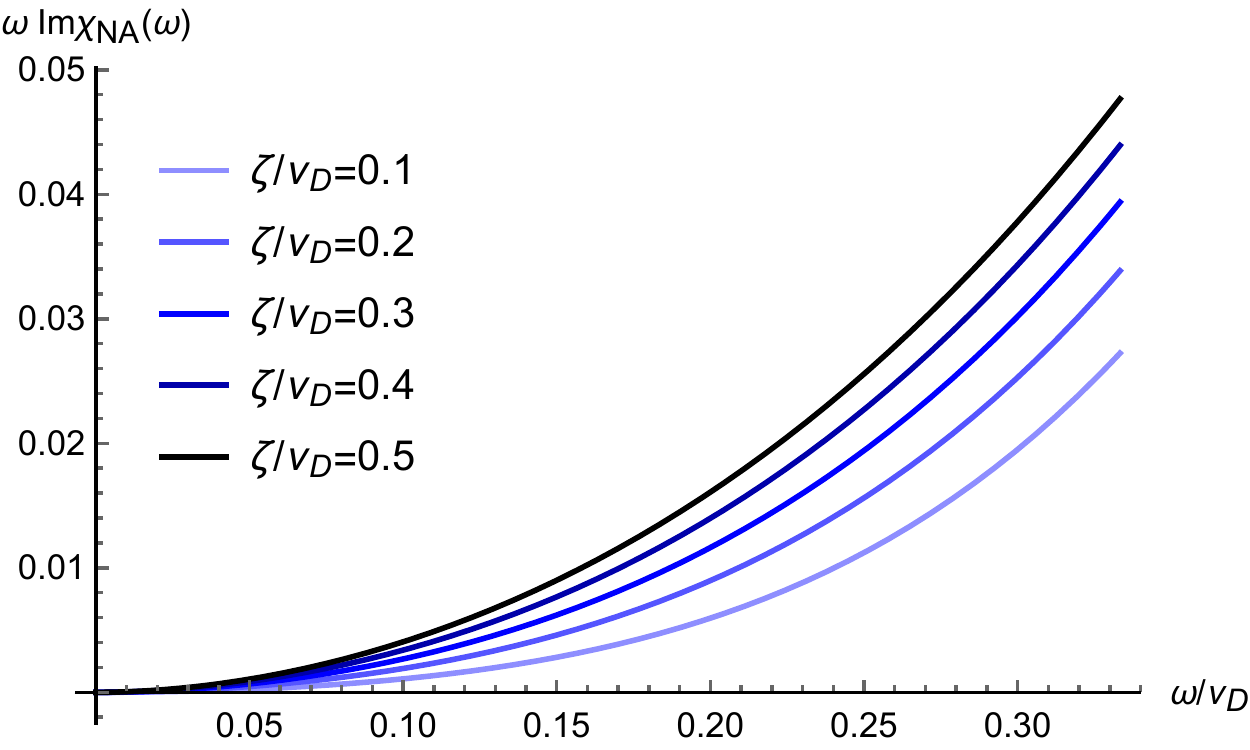}
     \caption{Real and imaginary parts of the nonaffine stress autocorrelation function at low frequency. Different colors represent different values of the microscopic damping $\zeta$. The dimensionful coefficient $c$ is set to unity.}
     \label{fig:1}
\end{figure}

For completeness, the low frequency behaviour of the nonaffine stress autocorrelation function is shown in Fig.\ref{fig:1} for different values of the damping $\zeta$. Interestingly, the correction to the shear modulus at finite (but low) frequency decreases with the damping $\zeta$, while the sound attenuation constant increases with it.\\

From our analytic expression, we can also derive, in the small damping limit $\zeta \ll 1$, the contribution from nonaffinity to the Rayleigh $\sim k^4$ term in the sound attenuation, the prefactor of which is given by:
\begin{equation}\label{stupenda}
    \Gamma_4=\frac{3\pi}{2}\,\frac{v^2\,\mu_{NA}}{m\,\nu_D^3}
\end{equation}
where $\mu_{NA}$ is the nonaffine part of the shear modulus and $m$ is the mass density. Interestingly, this contribution does not vanish in the limit $\zeta \rightarrow 0$ and it is therefore present even at $T=0$, as directly showed in the simulations of Ref.\cite{szamel2021sound}.\\

Finally, in the limit of small microscopic friction $\zeta$, our theory provides an analytic estimate of the crossover point $k_*$ between the quadratic and quartic scaling in \eqref{crosso} given by:
\begin{equation}\label{tesoro}
    k_*^2\,=\,\frac{2\,\nu_D\,\zeta}{\pi\,v^2}\,.
\end{equation}
As evident from this expression, upon increasing the temperature, the microscopic friction parameter $\zeta$ grows and the quadratic diffusive regime extends up to larger values of the wave-vector $k$. This behaviour is apparent in Fig.\ref{fig:2}.
\section{Discussion}
In summary, we presented an analytical theory of sound attenuation in amorphous solids starting from single particle motion and taking into account the inherently nonaffine dynamics. We derived in closed form the contributions from nonaffinity to both the hydrodynamic diffusive term $\sim \Gamma_2 k^2$ \eqref{mainresult} and the Rayleigh term $\sim\Gamma_4 k^4$ \eqref{stupenda}, with the prefactors $\Gamma_2, \Gamma_4$ expressed in explicit form as functions of important physical parameters. The Rayleigh attenuation due to nonaffine motions, \eqref{stupenda}, survives in the limit of zero temperature, or equivalently zero microscopic friction $\zeta=0$, and it has been recently shown via simulations to be the dominant process of sound attenuation in amorphous solids \cite{szamel2021sound}, while the harmonic disorder/HET contribution to the same scaling plays a comparatively very minor role. Furthermore, our theory analytically predicts the crossover from diffusive to Rayleigh damping at larger wavevector \eqref{tesoro}, which was trivially obtained before only from ad-hoc combinations of continuum HET and anharmonic theory, and neglecting the microscopic nonaffine dynamics \cite{Tomaras_Schirmacher}. Crucially, Refs. \cite{Tomaras_Schirmacher,Schirmacher_2006} are theories limited to the continuum level because their input are the fluctuations of the elastic moduli, whereas the micro-physics of the elastic moduli (i.e. how they are related to atomic-level displacements, bonding and atomic-level structure) is completely neglected.  Here, instead we have did consider the microphysics of the moduli, i.e. their full eigenmode decomposition into microscopic atomic vibrations, and, crucially the atomic nonaffine displacements which prove key to demonstrate the origin of sound attenuation at the microscopic level. \\

The analytical prediction of Rayleigh damping from nonaffinity is fully supported by recent numerical data~\cite{szamel2021sound}, which also prove its fundamental importance and dominance, with respect to the effects of harmonic disorder/HET, in the determination of sound attenuation in amorphous solids. Additionally, our main result in \eqref{mainresult}, namely the low-$k$ sound attenuation constant being quadratic in the frequency (or wave-vector) and increasing with the microscopic damping $\zeta$, is confirmed by numerical simulations in~\cite{D0SM02018D} where the internal friction is parametrized in terms of the inelasticity.

\begin{acknowledgments}
We thank Giorgio Frangi, Bingyu Cui and Dmitry Parshin for reading an early version of the manuscript and for providing useful comments. We thank Lijin Wang, Grzegorz Szamel and Elijah Flenner for correspondence and useful comments which helped us to improve our manuscript.
A.Z. acknowledges financial support from US Army Research Office, contract nr. W911NF-19-2-0055. M.B. acknowledges the support of the Shanghai Municipal Science and Technology Major Project (Grant No.2019SHZDZX01).
\end{acknowledgments}

\section*{ DATA AVAILABILITY}
The data that supports the findings of this study are available within the article.\\

\bibliography{refs}

\end{document}